\documentclass[12 pt,preprint]{aastex}
\shorttitle{3EG J2006-2321}
\shortauthors{Wallace et al.}
\begin{document}

\title{An AGN Identification for 3EG J2006-2321}

\author{P. M. Wallace}
\affil{Department of Physics, Astronomy, \& Geology, Berry College, Rome, GA 30149, USA\\ 
\rm{pwallace@berry.edu}}

\author{J. P. Halpern}
\affil{Columbia Astrophysics Laboratory, Columbia University, New York, NY
10027, USA\\ 
\rm{jules@astro.columbia.edu}}

\author{A. M. Magalh\~{a}es}
\affil{Instituto de Astronomia, Geof\'{\i}sica, e Ci\^{e}ncias Atmosf\'{e}ricas, \\
Universidade de S\~{a}o Paulo, S\~{a}o Paulo - SP 01060-970, BRAZIL\\ 
\rm{mario@iagusp.usp.br}}

\author{D. J. Thompson}
\affil{Code 661, NASA Goddard Space Flight Center, Greenbelt, MD 20771, USA\\
\rm{djt@egret.gsfc.nasa.gov}}

\begin{abstract}

We present a multiwavelength analysis of the high-energy gamma-ray source
3EG J2006-2321 ($l=18^{\circ}\!.82$, $b=-26^{\circ}\!.26$). The flux 
of this source above 100 MeV is shown to be variable on time scales of 
days and months. Optical observations and careful examination of archived 
radio data indicate that its most probable identification
is with PMN J2005--2310, a flat-spectrum radio quasar with a 5-GHz flux density
of 260 mJy. Study of the $V=19.3$ optical counterpart indicates a redshift
of 0.833 and variable linear polarization. No X-ray source has
been detected near the position of PMN J2005--2310, but an X-ray upper limit
is derived from ROSAT data. This upper limit provides for a
spectral energy distribution with global characteristics similar to those
of known gamma-ray blazars. Taken together, these data indicate that
3EG J2006--2321, listed as unidentified in the 3rd EGRET Catalog, is a
member of the blazar class of AGN. The 5-GHz radio flux density of this
blazar is the lowest of the 68 EGRET-detected AGN. The fact that
EGRET has detected such a source has implications for unidentified EGRET
sources, particularly those at high latitudes ($|b|>30^{\circ}$), many of
which may be blazars.

\end{abstract}


\section{Introduction}

Since the earliest days of gamma-ray astronomy, one of the foremost
questions in the field has been the identity of discrete sources.
In 1972-1973, SAS-2 was the first mission to detect radiation
from the Vela and Crab pulsars (Fichtel et al. 1975).
Launched in 1975, COS-B built on the
success of SAS-2 by detecting 25 point sources. However, only four of
those were identified (Swanenburg et al. 1981). As instruments became
more sophisticated, the number of identified and unidentified sources
has increased. The present list (Hartman et al. 1999, hereafter H99)
is comprised of the 271 sources detected by the Energetic Gamma-ray Experiment
Telescope (EGRET) on the late Compton Gamma-ray Observatory (CGRO) that
display significant flux above 100 MeV. Of these, 169 remain
unidentified. The 102 identified sources include a probable association
with the radio galaxy Cen A, a solar flare, and the LMC. The remaining
sources are pulsars (5), AGNs with low-confidence (27), and AGNs with
high-confidence
(66). The AGNs are blazars, typically flat-spectrum radio quasars (FSRQs) or
BL Lac objects. There is statistical evidence that
supernova remnants (Sturner \& Dermer 1995; Esposito et al. 1996), OB
associations (Kaaret \& Cottam 1996; Romero, Benaglia, \& Torres 1999),
and objects born in the Gould Belt (Gehrels et al. 2000) are also gamma-ray
emitters, but no single source type has been conclusively associated with any
of these classes.

Due to the relatively poor angular resolution of even the best gamma-ray
telescopes, efforts to identify sources depend on factors other than
spatial coincidence. Pulsars are typically found in the Galactic plane while
most blazars are found at high latitude where the diffuse Galactic emission
does not overwhelm the source photons. Additionally, variability over hours
and days offers evidence against identification as a pulsar (Ramanamurthy
et al. 1995) while blazars are known to have strongly variable gamma-ray
flux (e.g. Mattox et al. 1997a). But latitude and variability studies
cannot be used alone to identify individual gamma-ray sources with high
confidence as pulsars or blazars; for pulsars, a clear pulse profile must
be detected, while for blazars, multiwavelength studies are necessary as
these sources have a distinct broad-band signature. They are variable
at many frequencies, feature flat radio spectra, show variable polarization
at radio and optical frequencies, display power-law spectra at X-ray and
gamma-ray energies, and have moderate to large redshifts.

Recently there have been a number of papers published that describe
efforts to identify individual gamma-ray sources using multiwavelength
analyses (e.g., Mukherjee et al. 2000; Halpern et al. 2001; Mirabel
\& Halpern 2001; Reimer et al. 2001a). In the present paper a
multiwavelength study of 3EG J2006--2321 is presented, and the data
indicate that this source, listed as unidentified in H99, is a blazar.
In \S\S 2-5 the relevant data are presented; this is followed in
\S 6 by a short summary of the multifrequency data. The weak flux
density of the radio counterpart and its relevance to other
high-latitude unidentified EGRET sources is discussed in \S 7. In
\S 8 conclusions and suggestions for future work are summarized.

\section{Gamma-ray Observations}

The average flux displayed by 3EG J2006--2321 from 1991 April through
1995 September (the time span covered by H99) is
$(7.3\pm2.7)\times10^{-8}$ photons~cm$^{-2}$~s$^{-1}$ in the energy range
$E>100$ MeV, making it a relatively faint
EGRET source. The position of 3EG J2006--2321 is $l=18^{\circ}\!.82$,
$b=-26^{\circ}\!.26$ and the mean radius of the 95\% confidence contour is
about $0^{\circ}\!.67$. The source is well-isolated; its closest gamma-ray
neighbor is more than $8^\circ$ away and its high Galactic latitude
ensures that the Galactic diffuse radiation does not interfere with the
source photons. The EGRET spectrum (from 30 MeV to 30 GeV) is
consistent with a power-law photon index $\Gamma = 2.47\pm0.44$.

\begin{figure}[t!]
\epsscale{1.0}
\plotone{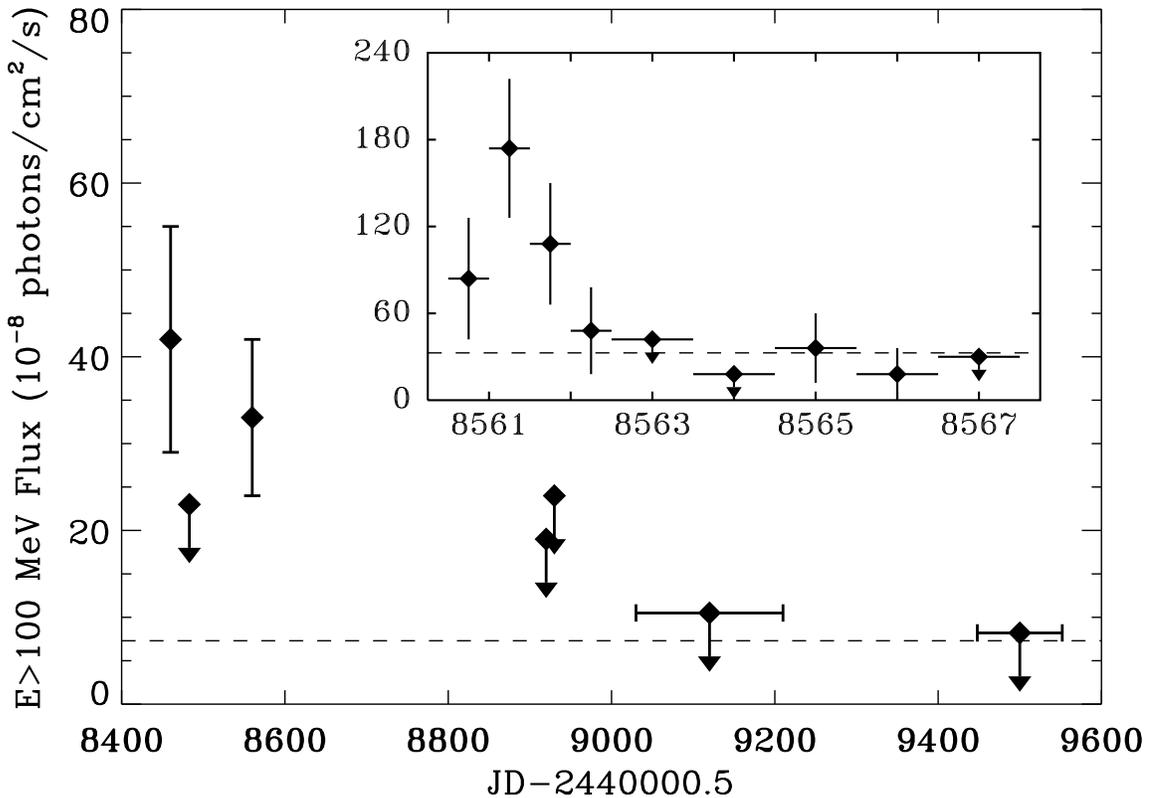}
\caption{Main plot: Flux history of 3EG J2006--2321 from 1991 April
to 1995 September, the time span covered by H99.
The first five points are averages over individual EGRET viewing
periods (VP's); each of the final two points represent averages over two
VP's; in no viewing periods after 1991 November
was the source detected. Inset plot: Light curve of 3EG J2006--2321
during VP 13.1, indicated by the third point on the main plot.
During this VP (1991 October 31-November 7) the source
was highly variable, with a peak flux nearly 24 times that of the average
over all VP's. The horizontal dashed lines indicate the mean flux for
the relevant time spans. \label{fig1}}
\end{figure}

A light curve of this source from CGRO Phase 1 through Cycle 4 is shown
in Figure 1. The first detection by any telescope of 3EG J2006--2321 occured during
EGRET Viewing Period (VP) 5.0 (1991 July 12-26) when it was detected
with a flux of $(44.1\pm12.7)\times10^{-8}$ photons~cm$^{-2}$~s$^{-1}$ at
a  significance of 4.4$\sigma$. During this time the source was
28$^{\circ}\!.$6 from the instrument axis, and there is no evidence that
the source was variable on a $\sim1$-day time scale during this
observation. Three weeks later 3EG J2006--2321
was 13$^{\circ}$6 from the EGRET axis and was not detected. From 1991
October 31-November 7 (EGRET VP 13.1) the source was again
13$^{\circ}\!.$6 off-axis and was detected with a flux of
$(32.7\pm8.7)\times10^{-8}$ photons~cm$^{-2}$~s$^{-1}$ at the 4.8$\sigma$
level. Although 3EG J2006--2321 was within 20$^{\circ}$ of the instrument
axis during six subsequent VP's, it was never again detected by EGRET.

There is evidence that this source is variable over time scales of weeks.
McLaughlin et al. (1996) describe a statistic, called the \it{variability index} 
\rm $V$. This statistic is used to calculate the probability that a given 
light curve is consistent with an intrinsically nonvariable source and is 
defined as follows. The $\chi^2$ for the light curve is
\begin{equation}
\chi^2=\sum_{i=1}^N \frac{(F-\bar{F}_i)^2}{\sigma_i^2}
\end{equation}
where $N$ is the number of observations, $F_i$ is the detected flux during the 
\it{i}\rm th observation, $\bar{F}$ is the mean flux for the viewing period, and
$\sigma_i$ is the 1 $\sigma$ flux uncertainty of the \it{i}\rm th observation. 
If $Q$ is the probability of obtaining a value of $\chi^2$ equal to or greater 
than the empirical $\chi^2$ from an intrinsically nonvariable source, then 
$V\equiv-\log Q$.  Roughly speaking, if $V<1$ then the source
in question is not considered to be variable; if $V>1$ then
the source may be variable. The variability index for the long-term 
light curve in Figure 1 is 2.9, corresponding to a probability of
$\sim0.0012$ that these data are produced by an intrinsically
nonvariable source.

During VP 13.1, the source was variable on a $\sim1$ day time scale.
The gamma-ray observations of 3EG J2006--2321 during this VP
are discussed in detail in Wallace et al. (2000); the relevant information
will be summarized here. A light curve is shown in the inset of Figure 1.
The first two days are broken down into four 12-hour periods;
the remaining points represent full days. The peak flux, centered in MJD
48560.25, is $(1.75\pm0.53)\times10^{-6}$ photons~cm$^{-2}$~s$^{-1}$; this 12-hour
detection has a significance of 5.4$\sigma$. The ratio of peak to average
flux for the VP is 5:1, and the ratio of peak flux to the overall EGRET
mean flux for 3EG J2006--2321 is nearly 24:1. It is very unlikely that the
source is not variable on this short time scale. Applying a $\chi^2$
test to the light curve yields a variability index of 3.2, corresponding
to a probability of 0.0006 that these data are produced by an intrinsically
nonvariable source.

\section{Radio Observations}

\begin{figure}[t!]
\epsscale{1.0}
\includegraphics[angle=90]{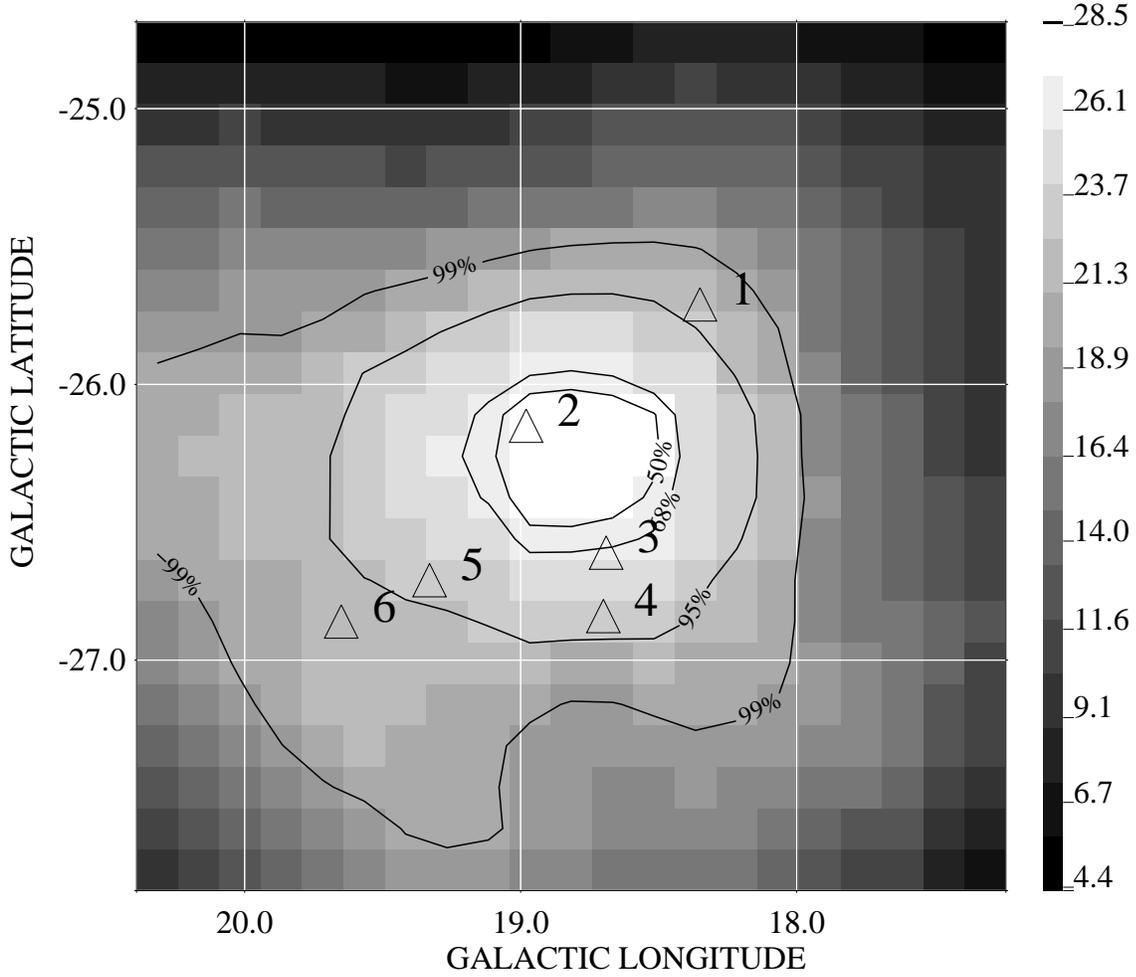}
\caption{Grey-scale sky map of 3EG J2006--2321.
The vertical scale on the right indicates the gamma-ray flux in units
of $10^{-8}$ photons~cm$^{-2}$~s$^{-1}$.
The positions of 5-GHz radio
sources from the PMN survey are marked with triangles and numbered
as in Table 1.
\label{fig2}}
\end{figure}

To find a radio counterpart to 3EG J2006--2321, we first search its error
circle for 5-GHz sources since gamma-ray blazars typically have
significant flux densities at this frequency (Mattox et al. 1997b).
At southern declinations, the Parkes-MIT-NRAO (PMN) survey
(Griffith \& Wright 1993) is the deepest at 5 GHz; we therefore look
to these sources. We expect that if 3EG J2006--2321 is indeed a blazar, it,
like other EGRET blazars, will be related to a flat-spectrum radio source,
with $-0.5\lesssim\alpha_r$ (where $F_{\nu}
\propto \nu^{\alpha_r}$). The NED database lists
six 5-GHz sources within the EGRET 99\% error contour; these sources
are listed in Table 1 with their 5-GHz flux densities, angular
separations from the gamma-ray position, and radio spectral indexes.
Figure 2 indicates their positions on a map of the gamma-ray intensity
of 3EG J2006--2321. These radio candidates are discussed below in order of
increasing right ascension.

\it{1. PMN J2003--2333.} \rm This source has a 5-GHz flux density ($S_5$)
of 43 mJy, the weakest of all the candidate sources, and it is $42^{\prime}$
from the EGRET position, between the 95\% and 99\% confidence contours.  
The nearest source to it in the
NRAO/VLA Sky Survey (NVSS; Condon et al. 1998) has a 1.4-GHz flux
density of 15.3 mJy, which, if identified with the PMN source
would give it a rising radio spectrum with an index
of 0.83.  However, the beam size of the Parkes telescope is
$4^{\prime}\!.2$, and there are at least four NVSS sources within
$3^{\prime}$ of the PMN position having a combined flux density of 45 mJy
at 1.4 GHz.  Given the weakness of all of these sources, and the
possibility of source confusion at 5 GHz, we do not consider this
to be a promising candidate.

\it{2. PMN J2005--2310.} \rm This source is the brightest of the six, with a
5-GHz flux density of 260 mJy. It is located well within the 50\%
confidence contour, at $11^{\prime}$ from the EGRET position. It appears
in the NVSS with a 1.4-GHz flux density of 302 mJy. We calculate
$\alpha_r=-0.12$ from these data. If we include the 365 MHz flux
density of 260 mJy from the Texas survey (Douglas et al. 1996) we
find that the radio spectrum is flatter, with $\alpha_r=0.002$.

\it{3. PMN J2007--2335.} \rm This source has a 5-GHz flux density of
141 mJy and is located near the 68\% confidence contour, at $23^{\prime}$
from the EGRET position. The nearest NVSS source has a flux density of 318 mJy,
giving it a spectral index of --0.65 between 1.4 and 4.85 (5)
GHz, below but near the threshold of -0.5 for flat sources. Its
lower-energy radio flux is relatively high; at 365 MHz it has a
flux density of 2.98 Jy. If this point is taken into account, its
spectral index falls to --1.2.  There is also the possibility of
some confusion with another NVSS source $4^{\prime}\!.7$ away
that has a flux density of 177.5 mJy.
However, in light of its relative
brightness and near-flat spectrum, we do not yet exclude
it from our analysis.

\it{4. PMN J2008--2338.} \rm This source has a 5-GHz flux density of 82
mJy and is $35^{\prime}\!.6$ from the EGRET position, just inside the 95\%
confidence contour. It has a 1.4-GHz counterpart with
flux density 275 mJy, giving it a steep spectrum with an index of
--0.97. It has a 993-mJy counterpart at 365 MHz; this does not
appreciably change $\alpha_r$. Only one EGRET blazar listed in H99, 
3EG J1937-1529, has a comparable radio spectral index (-0.96). 
Therefore the steepness of the radio spectrum is not unheard of 
among EGRET blazars, but it is rare. The combination of the steep 
radio spectrum, weak $S_5$, and large separation from the gamma-ray 
position leads us to exclude this source from consideration.

\it{5. PMN J2008--2305.} \rm This source has a 5-GHz flux density of 47
mJy and is $41^{\prime}$ from the EGRET position, just inside the 95\% confidence
contour. It has a 1.4-GHz counterpart with
flux density 41 mJy, giving it a flat spectrum with an index of +0.05.
There are no known counterparts at other frequencies.

\it{6. PMN J2009--2250.} \rm This source has a 5-GHz flux density of 49
mJy and is $55^{\prime}$ from the EGRET position, between the 95\% and 99\%
confidence contours. It has a 1.4-GHz counterpart
with flux density 97 mJy, giving it a spectral index of --0.55,
below but near the flat-spectrum threshold. If its 370-mJy
counterpart at 365 MHz is taken into account, the spectral index
falls to -0.78.

After excluding the two weak steep-spectrum sources, we are left
with PMN J2005--2310, PMN J2007--2335, PMN J2008--2305, and PMN
J2009--2250 as candidates for association with 3EG J2006--2310.
For gamma-ray blazars, the 5-GHz flux density has been found to correlate
linearly with peak gamma-ray flux at the 99.998\% confidence level
(Mattox et al. 1997b). (It should be noted that this correlation is not
well-understood and it serves only as a coarse guide. See \S 8.) There
are only ten EGRET blazars with peak flux above $10^{-6}$
photons~cm$^{-2}$~s$^{-1}$, and they all have $S_5>1.0$ Jy. We therefore expect to
find a relatively bright 5-GHz counterpart to 3EG J2006--2321 with its peak
flux of $(1.75\pm0.53)\times10^{-6}$ photons~cm$^{-2}$~s$^{-1}$.
However, the brightest of these candidates, PMN J2005--2310, has
$S_5=260$ mJy. Two of the other candidates are far dimmer still; the
flux densities of PMN J2008--2305 and PMN J2009--2250 are $>5$ times
weaker than that of PMN J2005--2310. Therefore if we are to
identify 3EG J2006--2321 with any of the four remaining radio sources, we
see that PMN J2005--2310 is the most compelling candidate, followed by
PMN J2007--2335. Another point in favor of these two sources is
that their angular separations from the EGRET position are smaller
than for the weaker sources. We therefore regard PMN J2008--2305
and PMN J2009--2250 as highly improbable candidates and restrict
ourselves to considering only PMN J2005--2310 and PMN J2007--2335.

\section{Optical Observations}

\subsection{PMN J2007--2335}

On 2001 June 29, six 300-s exposures of the field containing PMN
J2007--2335 were taken with a CCD on the 1.3m telescope of the
MDM Observatory. The combined $R$-band
image reveals a normal-looking galaxy at the position of the NVSS source,
(J2000) $20^{\rm h}07^{\rm m}25^{\rm s}\!.98,\
-23^{\circ}34^{\prime}35^{\prime\prime}\!.6$,
with $R\approx 19.8$. It is highly unlikely that such a galaxy is the
source of the high-energy gamma rays; therefore PMN J2007--2335 is
rejected as the counterpart of 3EG J2006--2321, leaving only
3EG J2006--2321 as a viable candidate.

\subsection{PMN J2005-2310}

Three CCD images of 200-s exposure
in the $V$ band centered on the position of PMN J2005--2310 were taken with
the MDM Observatory 2.4m telescope on 2000 July 24.
The central $80^{\prime\prime} \times 80^{\prime\prime}$
of the combined image is shown in Figure 3.
A point-like optical counterpart with $V=19.3$ was found at
(J2000) $20^{\rm h}05^{\rm m}56^{\rm s}\!.59,\
-23^{\circ}10^{\prime}27^{\prime\prime}\!.0$, within $1^{\prime\prime}$ of
the NVSS position of PMN J2005--2310, consistent with their
combined astrometric uncertainty.
An object at this position and comparable brightness can be seen on
various Digitized Sky Survey plates.
Galactic extinction is moderate at these latitudes; the absorption in the $V$
band is $\approx 0.5$ mag \citep{sch98},
giving the source an intrinsic $V=18.8$.

An optical spectrum of this object was obtained with the Goldcam
spectrometer on the 2.1m telescope of the Kitt Peak National Observatory (KPNO)
on 2000 June 3. 
A single 3000-s exposure using grating 240 and a slit
width of $1^{\prime\prime}\!.9$ yielded a resolution of $\approx 5$ \AA.
The target was acquired by blind offset from a bright star $10^{\prime\prime}$
to the east (see Figure 3), and the slit was oriented at the parallactic
angle, in this case N-S, which is essential at such a southerly declination
in order to obtain accurate spectrophotometry.
The fully reduced spectrum is shown in Figure 4.
It features a single broad emission line at 5129 \AA\ which,
from the absence of other emission lines over the observed spectral range,
we identify as Mg \small{II} \normalsize $\lambda$2798 at $z=0.833$.
The Mg \small{II} \normalsize line has the unique property that there is a 
range in redshift over which it is the only prominent quasar emission 
line that falls in the region observed by the typical optical spectrograph
in first order, which covers at most a factor of 2 in wavelength. Reliable 
redshifts are often achieved using Mg \small{II} \normalsize only. The neighboring broad emission
lines are [C \small{III}\normalsize] $\lambda$1909 and H$\gamma$ $\lambda$4340, 
which at $z=0.833$
would be redshifted to 3499\AA\ and 7955\AA, respectively, both outside of 
our observed spectral range. A narrow forbidden line that is sometimes seen
in quasars is [O~\small{II}\normalsize] $\lambda$3727, and there is a marginal detection of
this line at 6832\AA\ in our spectrum of PMN J2005--2310. The full width 
at half maximum of the broad Mg~II line is 3400 km s$^{-1}$, and its rest-frame 
equivalent width is 15.7 \AA, larger than that of BL Lac objects, but typical 
of FSRQs that have been identified with EGRET sources. The continuum flux 
is consistent with the $V$ magnitude measured from the CCD image obtained 
52 days later.

\begin{figure}[ht!]
\epsscale{0.6}
\plotone{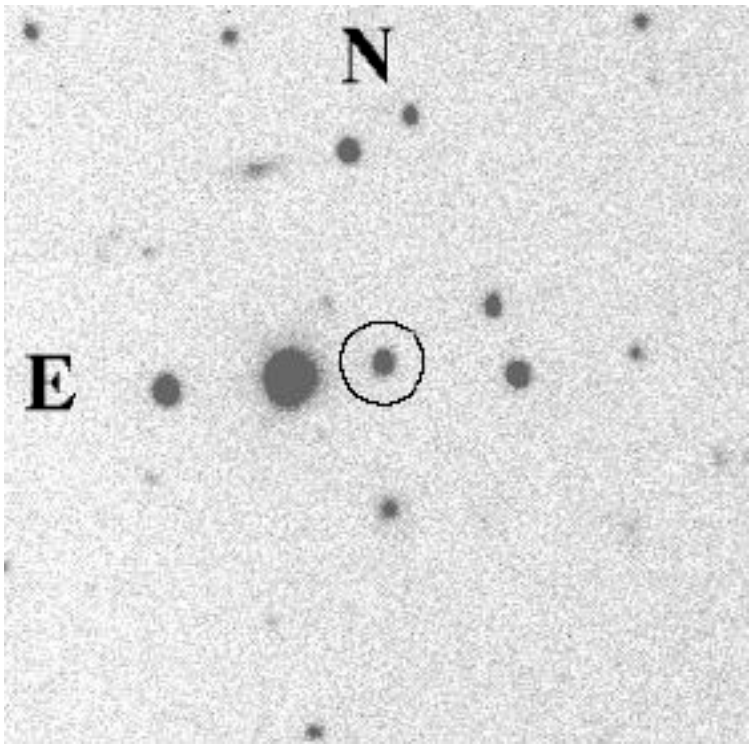}
\caption{Portion of a $V$-band image from the MDM 2.4m,
with the point-like optical counterpart of PMN J2005--2310 circled.
The field shown is $80^{\prime\prime} \times 80^{\prime\prime}$,
and the position of the $V = 19.3$ optical counterpart is
(J2000) $20^{\rm h}05^{\rm m}56^{\rm s}\!.59,\
-23^{\circ}10^{\prime}27^{\prime\prime}\!.0$, within $1^{\prime\prime}$
of the NVSS position. \label{fig3}}
\end{figure}

\begin{center}
\begin{figure}[ht!]
\epsscale{0.6}
\includegraphics[width=4in,angle=270]{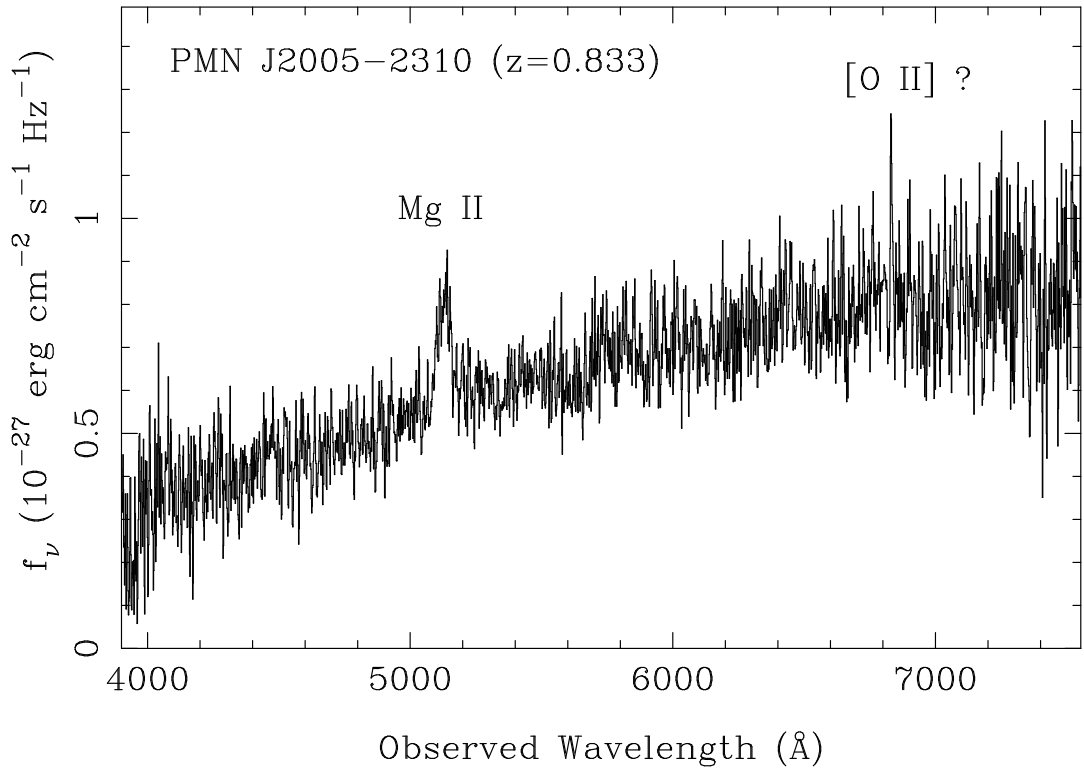}
\caption{Spectrum of the optical counterpart of PMN J2005--2310
from the KPNO 2.1m. The single broad emission line is identified as Mg \small{II} \normalsize $\lambda 2798$
and there is a possible detection of the forbidden line [O \small{II}\normalsize] $\lambda$3727.\label{fig4}}
\end{figure}
\end{center}

Strong and variable optical polarization is a characteristic blazar
signature \citep{pes97}, so optical polarimetry has been performed for
PMN J2005--2310. The source was observed with the IAGPOL
imaging polarimeter \citep{mag96} at the 61 cm IAG-USP telescope at
the Laborat\'orio Nacional de Astrof\'{\i}sca (LNA) on 2000 August~5
and on 2001 June~15. The polarimeter is a modification of the observatory's
direct CCD camera to allow for high-precision imaging polarimetry.
The first element in the beam is a 51mm diameter rotatable, achromatic
half--wave retarder followed by a Savart plate built by Opto
Eletr\^onica, S\~ao Paulo. One polarization modulation cycle is covered
for every $90\arcdeg$ rotation of the waveplate. The simultaneous
observations of the two beams allows observing under non-photometric
conditions at the same time that the sky polarization is practically
cancelled \citep{mag96}. Further details are given by \citet{kay99}.

The log of observations is presented in Table~\ref{imagepol}.
CCD exposures were taken through the $V$ filter with the waveplate
rotated through 16 positions (2000 Aug 5) and 12 positions (2001 June 15)
$22.5\arcdeg$ apart. The exposure time at each position was 300\,s.
The instrumental Stokes parameters $Q$ and $U$ were then obtained, as
well as the theoretical (i.e., photon noise) and measurement errors.
The latter are estimated from the residuals of the observations at each
waveplate position angle (${{\psi}_i}$) with regards to the
expected cos ($4{{\psi}_i}$) curve and are quoted in Table~\ref{imagepol};
they are consistent with the photon noise errors \citep{mag84}.

Due to the relative faintness (for the telescope) of PMN~J2005--2310,
care was taken to ensure that the estimates of sky values per pixel
were robust. We found that the {\it mode} option in the IRAF photometry
package gave consistent results for different annuli around the object.
The instrumental $Q$ and $U$ values were then
converted to the equatorial system from data of two polarized standard stars
\citep{tur90} obtained in each
night. The instrumental polarization was measured to be less than
0.03\% from observations of unpolarized stars in the same night. We
have hence applied no such correction to our data.

Table~\ref{imagepol} shows that on 2000 August 5, PMN~J2005--2310
was significantly polarized while that was not the case on 2001 June 15.
We conclude that PMN~J2005--2310 has a highly variable optical polarization,
a characteristic blazar signature, lending further credence to the association
of 3EG~J2006--2321 with PMN~J2005--2310.

The imaging polarimetry includes data on targets angularly close to
PMN~J2005--2310, thereby providing the means to estimate the
interstellar (IS) polarization towards that general direction. We used
a package, {\it pccdpack}, specially written for the analysis of field
stars \citep{per00}. A sample of 25 field stars in the 2000 images and
23 field stars in the 2001 images provided the weighted average polarization
values quoted in Table~\ref{imagepol}. The combined, weighted average for the
IS polarization is (1.47$\pm$0.01)\% @ 21.1$\arcdeg$. This
IS polarization estimate of about 1.5\% is entirely consistent
with the expected maximum percent IS polarization $P{_{max}} \le 9 E_{B-V}$
\citep{ser75}
and A$_V$=0.5 mag. Incidentally, this equatorial position angle
towards the direction of PMN~J2005-2310 corresponds to a Galactic position
angle $\theta_{G} = 91\arcdeg$, i.e., parallel to the Galactic plane,
confirming that we are measuring an IS component with the field stars.
The weighted IS polarization is much smaller than the 2000 August
observed polarization of PMN~J2005--2310 and hardly effects it; the intrinsic
polarization of the blazar on that date becomes (13.5$\pm$2.3)\% @
61$\arcdeg$.

\section{X-ray Observations}

The region surrounding 3EG J2006--2321 has had little exposure to X-ray
instruments. The sole data come from the ROSAT All-Sky Survey (Voges
et al. 1999; Voges et al. 2000) during 1990 October 6-26; in these data (both the Bright and
Faint Source Catalogs) there is no source within $20^{\prime}$ of the position
of PMN J2005--2310. This places an upper limit on the X-ray flux of the
source in question. A typical dim source in this region has a PSPC count
rate of $\approx 2.4\times10^{-2}$ counts s$^{-1}$.
Assuming a power-law spectrum with photon index
with $\Gamma=2.0$ and Galactic $N_{\rm H} = 8.5 \times 10^{20}$~cm$^{-2}$,
this corresponds to an unabsorbed flux between 0.1 keV and 2.0
keV of $\approx 7 \times 10^{-13}$ ergs~cm$^{-2}$~s$^{-1}$, which we
adopt as an upper limit for PMN J2005--2310.

\begin{figure}[h!]
\epsscale{1.0}
\plotone{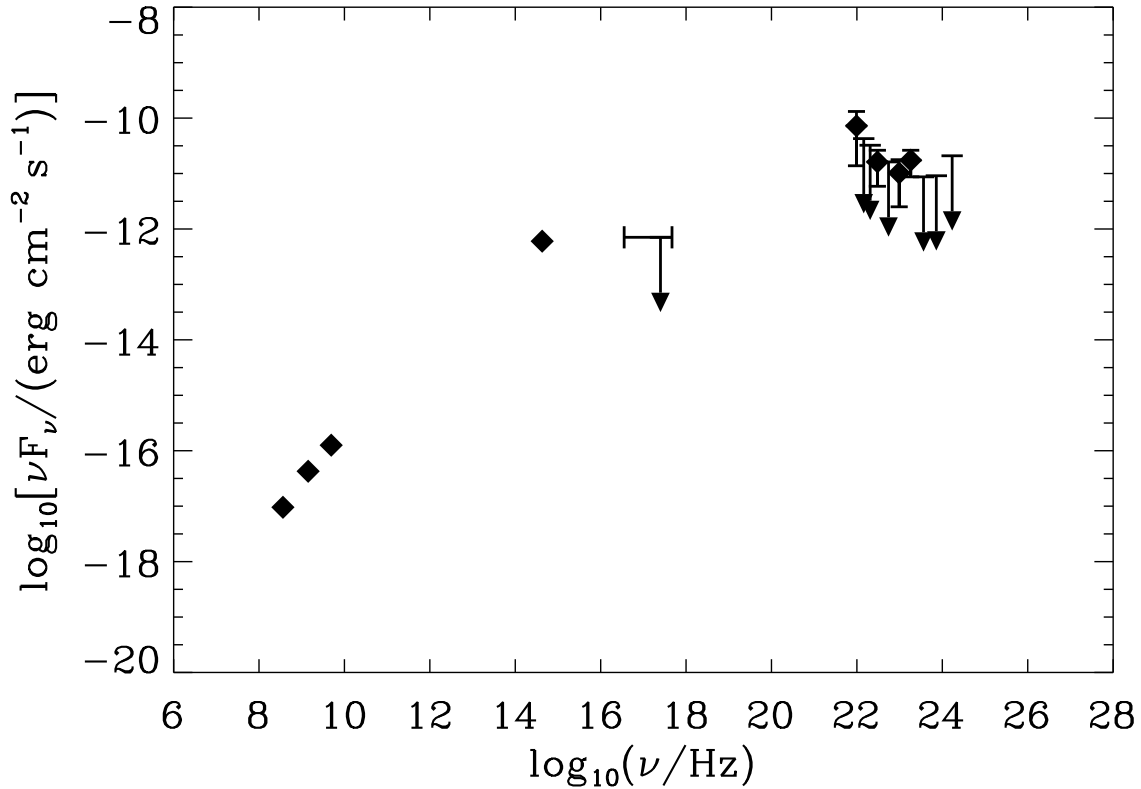}
\caption{Spectral energy distribution for 3EG J2006--2321. The error
bars of the radio and optical points fall within the diamond markers,
and markers with arrows indicate upper limits. The X-ray upper limit
is derived from ROSAT All-Sky Survey data and the gamma-ray limits are
statistical.
The data shown were not taken simultaneously. As the gamma-ray data
are representative of 3EG J2006--2321 in a high state, the gamma-ray
contribution to the bolometric energy flux is probably exaggerated
in this SED. \label{fig5}}
\end{figure}

\section{Summary of Multifrequency Data}

As 3EG J2006-2321 is faint, well off the Galactic plane, and has no
conspicuous objects of astronomical interest (e.g., SNR's, OB 
associations, X-ray sources) within its error circle, it
has had relatively little previous exposure to telescopes in any
frequency range. However, from the analysis of archived data in
the radio, X-ray, and gamma-ray regions, and from the results of
optical spectroscopy and polarimetry, it is evident that the source
is a member of the blazar class of AGN. Its gamma-ray flux is variable
on time scales of days and months. Its radio counterpart is a
flat-spectrum radio quasar (FSRQ) with a large redshift and strong,
variable optical polarization. Its X-ray upper limit makes its spectral
energy distribution (SED, Figure 5) consistent with a bimodal shape,
similar to SED's of known gamma-ray blazars (e.g., von Montigny
et al. 1995). Assuming a cosmology with $\Omega=1$ and $\Lambda=0$
and using the latest value of the Hubble constant $H_o=72$ km/s/Mpc
(Freedman et al. 2001), the peak gamma-ray isotropic
luminosity between 100 MeV and 10 GeV for 3EG J2006--2321
is $\sim4\times10^{47}$ ergs~s$^{-1}$, within the range of other flaring
gamma-ray AGN (e.g., McGlynn et al. 1997; Mattox et al. 1997a, 2001a).
Although there is evidence that the comological constant
is nonzero (Perlmutter et al. 1999), $\Lambda=0$ 
is assumed here only to normalize comparisons with previous 
calculations.

\section{On the Low Flux Density of PMN J2005--2310 and
         Implications for High-latitude Unidentified Sources}

It may be argued that the present source is unusual among gamma-ray
blazars as its 5-GHz flux density is weaker than that of any other
EGRET AGN and much weaker than any other blazar with comparable
peak gamma-ray flux. [The EGRET blazars with the lowest
5-GHz flux densities on record are 3EG J0743+5447 (272 mJy) and
3EG J2158--3023 (407 mJy). The weakest $S_5$ among EGRET blazars
with peak high-energy gamma-ray flux $>10^{-6}$ photons~cm$^{-2}$~s$^{-1}$
is 1080 mJy, from PKS 1406--076, associated with 3EG J1409--0745.]
However, $S_5$ of PMN J2005--2310 is not anomalous among radio
counterparts of gamma-ray blazars; it is merely low. Figure 6 shows
the distribution of 5-GHz flux densities of EGRET blazars;
PMN J2005--2310 is shown in black, other sources with peak gamma-ray
flux $>10^{-6}$ photons~cm$^{-2}$~s$^{-1}$ are gray. It can be seen that while
the radio flux density of the present source is weak, it conforms to
the prevailing distributions and does not alter them in any significant
way. Additionally, Mattox et al. (1997b), as was mentioned in \S 3, have
shown that for confirmed EGRET blazars, the 5-GHz flux density and
the peak gamma-ray flux are correlated. Mirabel et al. (2000) indicate 
that this correlation is not linear, but that there is a trend toward 
low $S_5/[F(>100)$ MeV] with increasing peak gamma-ray flux, which suggests 
that the identifications of Mattox et al. (2001b) are incomplete, 
and that low-flux radio sources should be counterparts of unidentified 
EGRET sources. The present source is in accordance with this trend;
it has the weakest 5-GHz flux density of any EGRET blazar, but has 
displayed the seventh-highest peak gamma-ray flux value. We conclude that, 
ultimately, 3EG J2006--2321 is not due any special attention for the weakness of 
its radio counterpart.

\begin{figure}[h!]
\epsscale{1.0}
\plotone{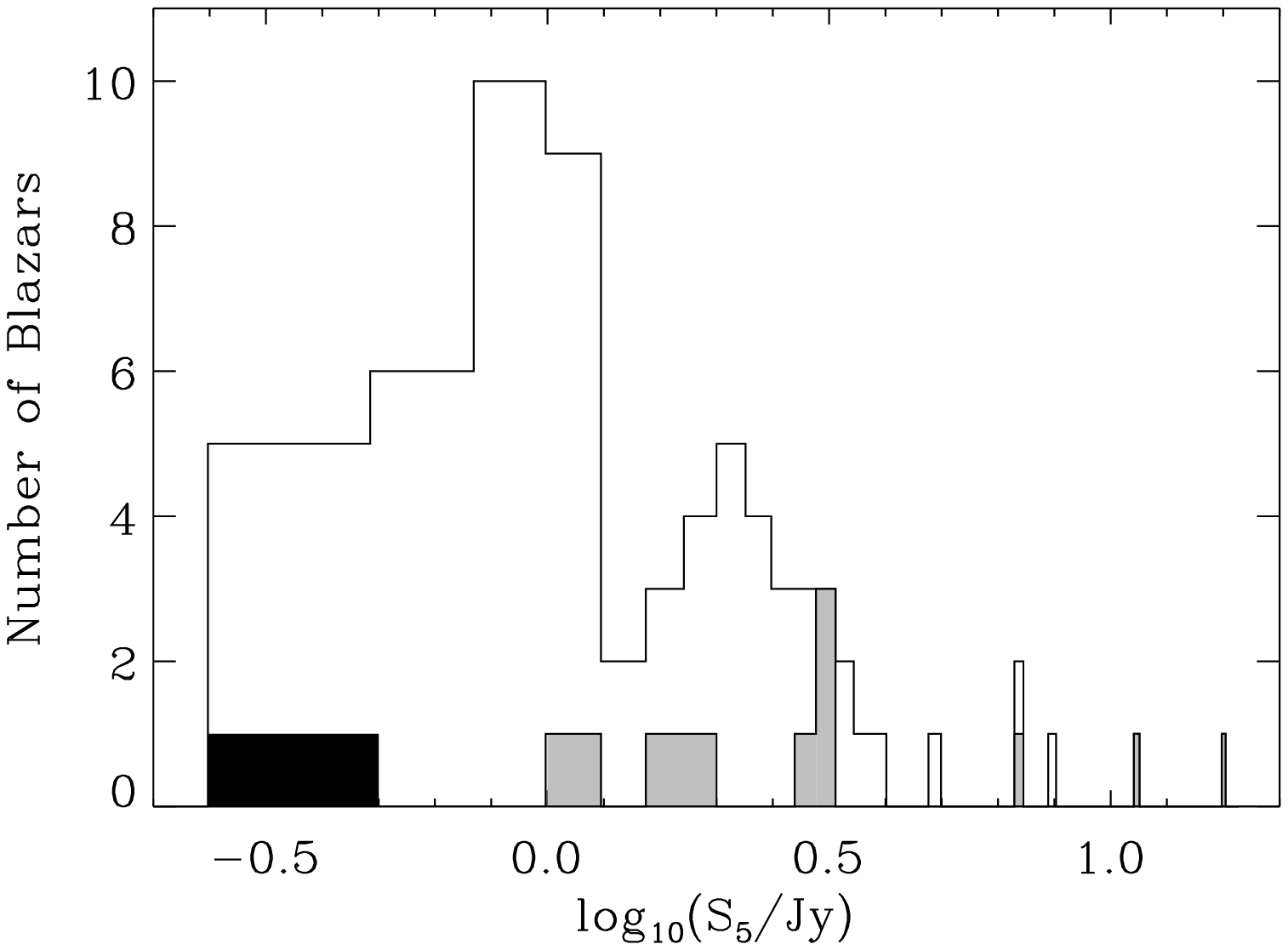}
\caption{Distribution of 5-GHz flux densities from EGRET blazars.
The bins are 0.25 Jy wide. The black rectangle represents 3EG J2006--2321
and the gray rectangles represent the 10 other blazars with peak
$E>100$ MeV flux $>10^{-6}$ photons~cm$^{-2}$~s$^{-1}$. There is an additional
source at 44.9 Jy; the peak $E>100$ MeV flux of this source is below
$10^{-6}$ photons~cm$^{-2}$~s$^{-1}$. The blazars used for this list are
the ``high-confidence'' AGN found in H99. The 2-Jy counterpart to
3EG J2016+3657 (Mukherjee et al. 2000) is also included. \label{fig6}}
\end{figure}

There are, however, implications of EGRET's ability to detect such a source.
Attention was first drawn to 3EG J2006--2321 when the EGRET database
was systematically searched for sources exhibiting
variability on short ($\sim1$ day) time scales (Wallace et al.
2000). All the 3EG sources were examined across all VP's from
1991 April to 1995 September, and 3EG J2006--2321 was the only
source to show new and compelling evidence of such variability.
All other instances of strong short-term variability uncovered by this
search had already been noted by investigators, connected with
previously-known AGN \footnote{The gamma-ray variability of 3EG
J0433+2910 led to its identification as a BL Lac object (Lundgren
et al. 1995; Wallace et al. 2001) but its flux variation was not on
a short time scale.}, and duly reported (e.g.,  Mukherjee et al. 1996;
Mattox et al. 1997a; McGlynn et al. 1997). The point to make about 3EG
J2006--2321 is not that its 5-GHz flux is so low, but that its
gamma-ray variability was strong enough for it to be picked out
by EGRET. There are probably other blazars with $S_5\lesssim500$
mJy near EGRET's threshold of detection, but no other happened to
flare so dramatically while in the EGRET field of view.
(Calculating the probability of detecting a flare
from such a near-threshold EGRET blazar is not practical.
Even if one assumed that all $|b|>30^{\circ}$ unidentified sources
are AGN whose variations were not detected in Wallace et al. (2000)
and that the population of known EGRET blazars is a good model,
the uncertainties inherent in such a calculation would render
the result meaningless.) Although a strongly variable
flux from a gamma-ray source indicates a probable blazar
identification, the reverse not true; that is, not all blazars
are indicated in the EGRET data by strongly variable flux
(McLaughlin et al. 1996). This point is especially relevant
for high-latitude sources, since they received, on average,
less exposure to EGRET than sources near the Galactic plane (H99),
and are less likely to be \it observed \rm to be variable.
We emphasize that the EGRET data do not rule out the possibility
that many unidentified sources with $|b|>30^{\circ}$ are blazars
similar to 3EG J2006--2321, with $S_5\lesssim500$ mJy.

Of course, there are other possibilities. There is evidence that
a population of dim unidentified sources at low and middle latitudes
is associated with the Gould Belt (Gehrels et al. 2000), which reaches
a maximum of 30$^{\circ}$ above the Galactic center. There may be
some ``stragglers'' from the Belt at latitudes higher than this, but
even if this is the case the number of such sources must be small.
Galaxy clusters may be responsible for a very small number of
unidentified EGRET sources at high latitudes (Colafrancesco et al. 2001)
but any high-energy gamma-ray emission from such clusters
probably falls below EGRET's sensitivity (Reimer \& Sreekumar 2001b).
It is also possible that the Galactic halo houses
some of the persistent EGRET sources (Grenier et al. 2001), but to date
this is only a conjecture. It remains that the only identified
gamma-ray sources above 30 degrees Galactic latitude are blazars
with flat-spectrum radio counterparts, and we suggest that
there are others among the 30 unidentified sources with $|b|>30^{\circ}$.

The present case brings to mind weaknesses in standard identification
methods that prevent these blazars from being identified. These
weaknesses cannot be easily overcome, as they are manifestations
of insufficient data and theory; however, it is important
to be reminded of them. It is common practice (Thompson et al. 1995; H99)
to use 5-GHz sources in the selection of radio counterparts.
Specifically, radio sources that are loud ($\gtrsim1$ Jy) and flat
($-0.5\lesssim\alpha_r$) at this frequency have been
considered to be top candidates for association with EGRET AGN.
There are two weaknesses to this approach worth mentioning here.
First, \it{in general} \rm it does not easily work for dim
($S_5\lesssim500$ mJy) radio sources, because the sky density
of such sources is high and source confusion becomes a problem,
especially for weak EGRET sources with large error circles.
In the present work we were forced to look at weak 5-GHz sources
that were not considered in the original identification process;
because of the low number of 5-GHz candidates we are able to
determine the appropriate counterpart. Also, as Bloom \&
Dale (2001) have noted, the use of this frequency to establish
the loud, flat-spectrum nature of radio counterparts to gamma-ray
AGN has no truly compelling physical justification but
is a necessity; there are no complete radio surveys at higher
frequencies with sensitivity below $\sim1$ Jy. Indeed, 3EG J0743+5447
($S_5=272$ mJy) is representative of a small group of EGRET blazars
that are fairly dim and fairly flat at 5 GHz but have been found to
have brighter and flatter spectra extending beyond 200 GHz (Bloom et
al. 1997); 3EG J2006--2321 may be similar.

Other, more sophisticated means of identifying EGRET sources with
flat-spectrum radio sources have been tried. In the most recent of these,
Mattox, Hartman, \& Reimer (2001b) searched for potential
radio counterparts to all
sources listed in H99, allowing for sources with arbitrarily low $S_5$.
Their approach is a quantitative one; for a given radio source and a
given EGRET source the probability of association is calculated, taking
into account certain properties of the radio source. Among these properties
are the 5-GHz flux density, the spectral index (when available) near 5
GHz, the angular separation from the EGRET position, and the
sky density of sources at least as bright and flat as the one in question.
The researchers list 46 blazar identifications with a ``high probability''
of being correct, and 37 additional radio associations that are
considered to be ``plausible''. The former list includes no unidentified
sources and the latter includes 15 unidentified  sources, none of which
are in the region $|b|>30^{\circ}$. This does not mean that there are no
blazars among this group; by their generic calculation 3EG J2006--2321
does not meet the criteria for either list.\footnote{However, 3EG
J2006--2321 is included on the ``plausible'' list in light of its gamma-ray
variability, which is not factored into their standard calculation.} A
radio source must have a $\sim0.04$ probability of association
with the gamma-ray source to be included among the plausible associations.
The probability of association of PMN J2005--2310 with 3EG J2006--2321
was calculated to be 0.015, and the four ``high-probability'' blazars with
$S_5<1$ Jy have more than 10 times the probability of being associated
with their radio counterparts than does 3EG J2006--2321 with PMN J2005--2310.
As mentioned above, high-latitude sources are the most likely to be
indentifiable as blazars; it is worth noting that if any of these 30
sources are blazars, and it seems likely that \it some \rm should be,
the method of Mattox et al. (2001b) is not helpful for indicating them.
Stated another way, although this method may not prove that a particular 
blazar is the true identification with high confidence, that does not mean 
that such an identification is false with high confidence.

The question should be asked: is it possible to make associations
of non-variable EGRET sources with weak 5-GHz radio counterparts?
If so, new search methods must be used. High-latitude sources
are the most likely to be identified as blazars and have
the easiest error circles to search because in the region
$|b|>30^{\circ}$ the sky densities of gamma-ray sources and
potential radio counterparts are low, and because in this region the
diffuse Galactic gamma radiation does not wash out source photons. Careful
searches for blazars within high-latitude EGRET error circles may be
helpful for pointing out likely identifications, and cases
in which no blazar can be found within the error circle would be interesting. 
Such work has already begun; Bloom \& Dale (2001) have searched the
error circles of some high-latitude unidentified sources and are in
the process of monitoring optical counterparts of relevant flat-spectrum
radio sources. Their candidate radio counterparts have 5-GHz
flux densities between 33 mJy and 440 mJy. Optical spectroscopy
is planned for some of these sources. Similar work on other sources
is encouraged. The present work should embolden such investigations
as it provides additional evidence that EGRET has detected blazars with
radio properties similar to those for which they are searching.

\section{Conclusion}

From analysis of archived radio, X-ray, and gamma-ray data and our own 
optical spectroscopy and polarimetry, we conclude that 3EG J2006-2321
is a member of the blazar class of AGN.
This identification is interesting because it is a reminder that EGRET
is capable of detecting blazars with $S_5$ on the order of a fourth of
a Jansky. The remaining EGRET unidentified sources most likely to be
identified as blazars are in the region $|b|>30^{\circ}$. 
Further searches for possible radio and optical counterparts within
the error circles of the 30 unidentified sources in this region are
encouraged.

While the present analysis is sufficient to identify 3EG J2006--2321,
little more can be said about the source, and no conclusions are reached
regarding beaming and radiation mechanisms of gamma-ray-bright AGN.
Complete and simultaneous multiwavelength observations are needed
to constrain blazar models. The Gamma-ray Large Area Space Telescope
(GLAST), scheduled for launch in 2006, is expected to uncover thousands
of gamma-ray blazars and other high-energy sources; however, in order
to realize a full return of GLAST science, these sources must be
observed not just in gamma rays, but across the electromagnetic spectrum.

\acknowledgments

PMW gratefully acknowledges support from the NASA/ASEE Summer Faculty
Fellowship Program and from an AAS/NASA Small Research Grant (2000).
The authors thank O. Reimer for his helpful comments and sustained interest 
in this project. AMM is thankful for support from S\~ao Paulo State 
funding agency FAPESP (97/11299-2) and CNPq, and is grateful to Rocio 
Melgarejo and Fernando Nascimento da Silva for help with data gathering 
and to Antonio Pereyra for help with \it{pccdpack}. \rm This research made 
use of the NASA/IPAC Extragalactic Database (NED) which is operated by 
the Jet Propulsion Laboratory/California Institute of Technology under 
contract with NASA.

\begin{center}
\begin{deluxetable}{ccccc}
\footnotesize
\tablecaption{5-GHz Sources Within the 3EG J2006--2321 99\% Confidence Contour.
\label{tbl-1}}
\tablewidth{0pt}
\tablehead{\colhead{Source No.}&
\colhead{PMN Coordinate Name} &\colhead{$S_5$ (mJy)} & \colhead{Angular Separation
(arcmin)} & \colhead{$\alpha_r\tablenotemark{a}$}
}
\startdata
1. & J2003--2333 & $43\pm11$  & 41.7 & ...    \\
2. & J2005--2310 & $260\pm17$ & 11.0 & --0.12 \\
3. & J2007--2335 & $141\pm13$ & 22.9 & --0.65 \\
4. & J2008--2338 & $82\pm11$  & 35.6 & --0.97 \\
5. & J2008--2305 & $47\pm11$  & 40.8 & +0.05  \\
6. & J2009--2250 & $49\pm11$  & 55.3 & --0.55 \\
\enddata
\tablenotetext{a}{The radio spectral index $\alpha_r$ (where $F_{\nu}
\propto \nu^{\alpha_r}$) is calculated from the flux
densities at 1.4 and 4.85 (5) GHz}
\end{deluxetable}
\end{center}

\begin{deluxetable}{llcccc}
\tablecaption{ PMN~J2005--2310 Field Imaging Polarimetry Observations}
\tablewidth{0pt}
\tablehead{
        \colhead{Object}   &
        \colhead{Date} &
        \colhead{Exp} &
        \colhead{P}   &
        \colhead{$\sigma_P$}  &
        \colhead{$\theta$}
        \\[.2ex]
        & \colhead{(UT)} &
        \colhead{(min)} & \colhead{\%} & \colhead{\%} &
        \colhead{(deg)} 
        }
\startdata
PMN~J2005--2310 & 2000 Aug 5 & $16\times5$ & 13.8 & 2.3 & 58 \\
& 2001 Jun 15 & $12\times5$ & 4.8 & 3.4 & 35 \\
Field Stars & 2000 Aug 5 & $16\times5$ & 1.465 & 0.013 & 20.4 \\
& 2001 Jun 15 & $12\times5$ & 1.492 & 0.016 & 22.1 \\
\enddata
\label{imagepol}
\end{deluxetable}

\end{document}